\documentclass[twocolumn,superscriptaddress,floatfix,preprintnumbers,prl,hyperref,nofootinbib]{revtex4-2}
\usepackage[colorlinks=true,breaklinks=true]{hyperref}
\usepackage{comment}
\usepackage{physics}
\usepackage[utf8]{inputenc}
\hypersetup{allcolors=[rgb]{0.0 0.0 0.6},linkcolor=[rgb]{0.75 0.05 0.05}}
\usepackage{mathtools}
\usepackage[dvipsnames]{xcolor}
\usepackage{float}
\usepackage{letltxmacro}
\LetLtxMacro{\oldcite}{\cite}

\renewcommand{\cite}[1]{\mbox{\oldcite{#1}}}

\usepackage{orcidlink}
\usepackage{enumitem}

\long\def\exclude#1{}

\newcommand{\bn}{{\bf n}}
\newcommand{\bp}{{\bf p}}
\newcommand{\bk}{{\bf k}}
\newcommand{\bD}{{\bf D}}
\newcommand{\br}{{\bf r}}
\newcommand{\bK}{{\bf K}}
\newcommand{\bv}{{\bf v}}
\newcommand{\GF}{G_{\rm F}}

\begin{document}

\title{Neutrino-Mass-Driven Instabilities as the Earliest Flavor Conversion in Supernovae}

\author{Damiano F.\ G.\ Fiorillo \orcidlink{0000-0003-4927-9850}}
\email{damianofg@gmail.com}
\affiliation{Deutsches Elektronen-Synchrotron DESY,
Platanenallee 6, 15738 Zeuthen, Germany}

\author{\hbox{Hans-Thomas Janka \orcidlink{0000-0002-0831-3330}}}
\affiliation{Max-Planck-Institut f\"ur Astrophysik, Karl-Schwarzschild-Str.~1, 85748 Garching, Germany}

\author{Georg G.\ Raffelt \orcidlink{0000-0002-0199-9560}}
\affiliation{Max-Planck-Institut f\"ur Physik, 
Boltzmannstr.~8,  85748 Garching, Germany}

\begin{abstract}
Collective neutrino flavor conversions in core-collapse supernovae (SNe) begin with instabilities, initially triggered when the dominant $\nu_e$ outflow concurs  with a small 
antineutrino flux of
opposite lepton number, with $\overline{\nu}_e$ dominating over $\overline{\nu}_\mu$. When these ``flipped'' neutrinos emerge in the energy-integrated angular distribution (angular crossing), they initiate a fast instability. However, before such conditions arise, spectral crossings typically appear within $20~\mathrm{ms}$ of collapse, i.e., local spectral excesses of $\overline{\nu}_e$ over $\overline{\nu}_\mu$ along some direction. Therefore, post-processing SN simulations cannot consistently capture later fast instabilities because the early slow ones have already altered the conditions.
\end{abstract}

\maketitle

{\bf\textit{Introduction.}}---At the end of its life, the core of a massive star collapses and forms a protoneutron star (PNS). It releases an intense flux of neutrinos, which likely is the primary driver of the subsequent supernova (SN) explosion~\cite{Colgate:1966ax,Bethe:1985sox,Janka:2012wk}. Neutrinos depositing energy in the gain region revive the stalled shock wave, thereby triggering the delayed explosion. They also regulate nucleosynthesis in the ejecta. Neutrino interactions, which strongly depend on flavor, are thus key microphysical ingredients in these spectacular cosmic fireworks.

Intriguingly, when neutrinos are dense, their flavor evolution famously
remains an open challenge~\cite{Tamborra:2020cul, Volpe:2023met, Johns:2025mlm}. \hbox{Collective} flavor conversions (CFCs) dominate---flavor exchange among neutrinos driven by the coherent weak field they themselves generate. This process is akin to energy exchange between electrons via electric fields in a plasma. Though typically negligible, this weak field can grow exponentially, representing a flavor instability, which occurs when the dominant $\nu_e$ outflow is accompanied by a small excess in the energy and angular distribution of opposite lepton number (flipped neutrinos~\cite{Fiorillo:2024bzm}). Such instabilities may develop below the shock wave and could alter energy deposition and the explosion dynamics~\cite{Nagakura:2023mhr, Ehring:2023lcd, Ehring:2023abs, Wang:2025nii}.

Instability is a diagnostic tool: SN simulations neglect flavor conversions, so instabilities reveal inconsistency of the subsequent evolution. Johns has stressed the corresponding inconsistency in mapping a pre- to post-instability configuration~\cite{Johns:2024dbe}. When an instability arises, it disrupts the evolution, rendering later ones unphysical. The only meaningful concept is that of weak instabilities that emerge gradually from a previously stable configuration \cite{Fiorillo:2024bzm, Fiorillo:2024uki}. Most attention has focused on fast instabilities~\cite{Izaguirre:2016gsx}, which grow within tens of picoseconds and thus appear particularly disruptive. However, not the fastest instabilities are most significant, but rather the first ones to appear, as these  will shape the subsequent evolution. This is the key message of this Letter.

\begin{figure}
    \centering
    \includegraphics[width=1.0\columnwidth]{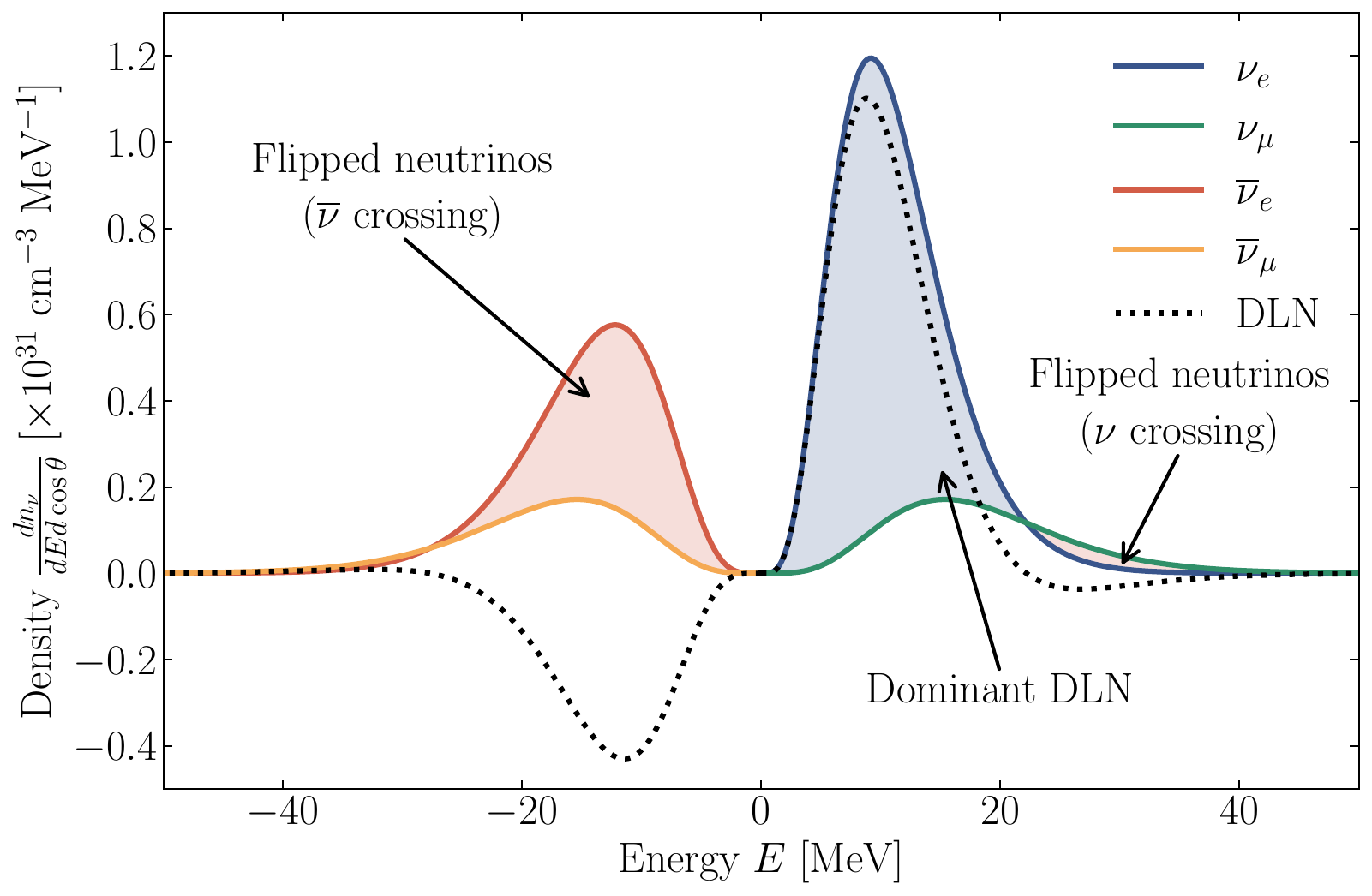}
    \caption{Double-crossed spectrum from our SN model at $62\,\mathrm{ms}$ post bounce, $r = 100\,\mathrm{km}$, and an angle to the radial direction of $\cos\theta = 0.6$  (asterisk in Fig.~\ref{fig:flipped}). Spectra reconstructed from discretized energy bins using a Gamma fit, as in Ref.~\cite{Tamborra:2012ac}. The total (i.e.~energy-integrated) DLN is positive; the positive contribution is highlighted with blue shade. A local excess in an energy interval of flipped neutrinos, with the opposite DLN (red shade), forms a spectral crossing and can cause an instability. Dotted line for DLN energy distribution.
    }
    \label{fig:crossing}
\end{figure}

Fast instabilities require angular crossings (the distribution flipping from $\nu_e$ to $\overline{\nu}_e$ dominance), which may or may not occur in 1D simulations~\cite{Tamborra:2017ubu, Shalgar:2019kzy, Morinaga:2019wsv, Capozzi:2020syn, Cornelius:2025nvd}, while they seem generic after some 100~ms in multi-D models~\cite{Abbar:2018shq,Nagakura:2019fas,DelfanAzari:2019tez,Abbar:2019zoq,Glas:2019ijo,Abbar:2020qpi,Nagakura:2021hyb}. In contrast, ``slow'' instabilities, those induced by neutrino masses, require only a spectral crossing: dominance in some energy interval of $\overline{\nu}_e$ over $\overline{\nu}_\mu$ ($\overline{\nu}$ crossing) or of $\nu_\mu$ over $\nu_e$ ($\nu$ crossing). Such CFCs consist of pairwise, collisionless conversions $\overline{\nu}_e\nu_e\to \overline{\nu}_\mu \nu_\mu$ or $\nu_e \nu_\mu\to \nu_\mu \nu_e$, reshuffling lepton number across the energy distribution. They are triggered by a sign change of the difference of $\nu_e$ and $\nu_\mu$ lepton number (DLN), defined as $n_{\rm DLN} = n_{\nu_e} - n_{\nu_\mu}$ for neutrinos and $n_{\rm DLN} = n_{\overline{\nu}_\mu} - n_{\overline{\nu}_e}$ for antineutrinos.

Figure~\ref{fig:crossing} shows a realistic double-crossed spectrum that we will use as a reference case. We show antineutrinos as neutrinos with negative energy, and use $n_\nu$ to denote number densities. The dominant DLN is positive; the plasma is dominated by $\nu_e$. There is a ``flipped'' range in both the $\nu$ and $\overline\nu$ sector, where the DLN is negative, i.e., opposite to the dominant one. 

In this Letter, we show that $\overline{\nu}$ crossings emerge immediately after collapse, making slow instabilities the first to appear. Despite their name, they grow quickly enough to alter the flavor distribution well before angular crossings develop, undermining any meaningful search for fast instabilities at later times, when the simulation has already become fundamentally inconsistent. 

The mechanism by which these early slow instabilities arise is generic: after the neutronization burst, the $\overline{\nu}_e$ overshoot their heavy-flavor partners and cause a spectral crossing. We illustrate this development in a post-processed SN simulation, exhibiting the progressive breakdown of the consistency of post-processing, with stronger neutrino-mass-driven instabilities growing within tens of ms. This early phase of SN evolution does not depend on detailed model assumptions or 3D effects, so our specific SN model is representative for the generic effect. We also stress that the wavenumbers and growth rates of the slow instabilities follow from our recent analytic estimates \cite{Fiorillo:2025zio}---they could not have been identified in a global numerical treatment.

{\bf\textit{Flavor instabilities.}}---Flavor instabilities emerge in SNe when, relative to the dominant $\nu_e$ flux, a small population carries the ``wrong''---opposite to the dominant one---lepton number. This role is played, in our two-flavor setup, by $\overline{\nu}_e$ or $\nu_\mu$. Flavor dynamics tries to avoid species that are wrong in this sense, so these subdominant species flip identity by emitting a flavomon---the quantum of the weak field they generate---through the decays $\overline{\nu}_e \to \overline{\nu}_\mu + \psi$ and $\nu_\mu \to \nu_e + \psi$. If there is an energy interval in which $\overline{\nu}_e$ dominate over $\overline{\nu}_\mu$, their decays dominate over the inverse reactions, causing an instability; the same can happen if there is an energy interval with $\nu_\mu$ dominating over $\nu_e$. The lepton number extracted from the flipped neutrinos is not lost, but converted to collective flavor waves and exchanged with the much larger $\nu_e$ flux. As more flavomons are emitted, the field grows, representing CFC. A full account of this mechanism is provided in the Supplemental Material (SM)~\cite{supplementalmaterial}.

Spectral $\overline{\nu}$ crossings are unavoidable in SN evolution, driven by the increasing $\overline{\nu}_e$ flux relative to $\overline{\nu}_\mu$ after core bounce. In addition, $\nu$ crossings may also occur, owing to the harder $\nu_\mu$ spectra compared to $\nu_e$. To assess whether ``slow'' CFCs can significantly affect the subsequent evolution, we must go beyond this qualitative picture and quantify when and where the crossings emerge, and how rapidly the resulting instabilities grow.

The answer requires solving the dispersion relation for collective modes, which arises from the quantum kinetic equation~\cite{Dolgov:1980cq, Rudsky, Sigl:1993ctk, Fiorillo:2024fnl, Fiorillo:2024wej}, linearized under the assumption of small flavomon field amplitude. The field evolves as a superposition of waves with wavevectors $\bK$---assumed to satisfy $|\bK| \gg \ell^{-1}$, where $\ell$ is the characteristic scale of SN inhomogeneities---each oscillating with frequency $\Omega_\bK$ and possibly growing at a rate $\gamma_\bK$.

Treating each $\bK$ mode separately furnishes the dispersion relation in the form of an integral equation, which is surprisingly cumbersome to solve in practice. One needs to rely on root-finding algorithms in the manner of a black-box approach that obscures physical insight. We deliberately make the opposite choice and rely on the approximations developed in Refs.~\cite{Fiorillo:2024bzm, Fiorillo:2024uki, Fiorillo:2024pns, Fiorillo:2025ank, Fiorillo:2025zio}, which are derived from first principles and validated by comparison with SN-inspired numerical solutions. These formulas relate $\gamma_\bK$ to the DLN parameters, providing a physical order-of-magnitude strategy.

\begin{figure*}
\includegraphics[width=\textwidth]{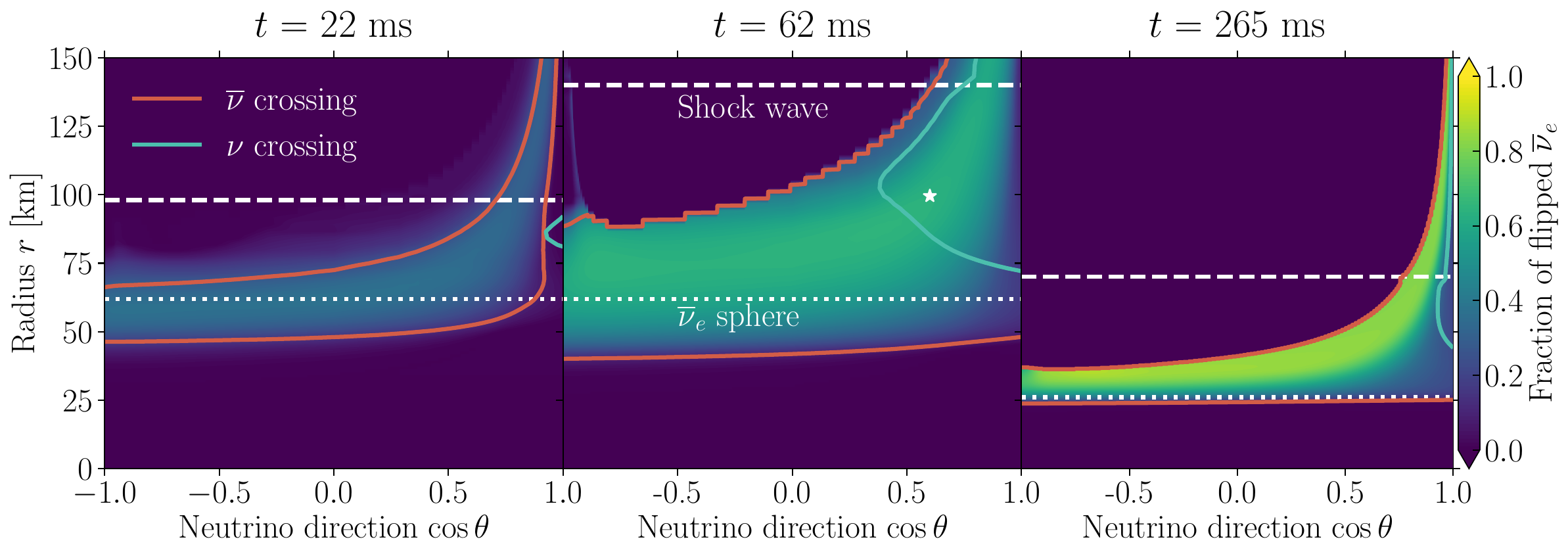}
\caption{Fraction of flipped $\overline{\nu}_e$ color-coded as a function of radius and direction at different times. Regions exhibiting a $\overline{\nu}$ or $\nu$ crossing are indicated by colored lines, using a threshold of 5\% for the flipped fraction. The shock wave radius is shown (dashed white line) and the $\overline{\nu}_e$ sphere (dotted white), defined as the radius where the $\overline{\nu}_e$ number flux equals one-quarter of the density. The number-changing reactions for $\overline{\nu}_\mu$ freeze out much deeper inside. The asterisk marks the location for the energy distribution shown in Fig.~\ref{fig:crossing}.}\label{fig:flipped}
\end{figure*}

Slow instabilities can display qualitatively distinct forms of behavior, depending on the underlying neutrino distribution. We identify three classes according to the depth of the spectral crossing, and elaborate in more detail in the SM~\cite{supplementalmaterial}. This subdivision is not redundant: the differences 
between these classes are the same as those between the bump-on-tail and beam-plasma instabilities in ordinary plasmas, emerging continuously from one another, yet with distinct physical properties \cite{boyd2003physics}.

For very shallow spectral crossings, we have \textit{narrow slow instabilities}~\cite{Fiorillo:2025zio}, due to the narrow resonance in the neutrino-flavomon interaction. These are \textit{directional}, with flavomons emitted collinearly with the parent neutrinos.

The order of magnitude of $\gamma_\bK$ is proportional to the energy splitting of neutrinos with opposite lepton number and energy $E$, which is $\omega_E=\delta m^2 \cos 2\vartheta/2E$, with $\delta m^2$ the squared-mass splitting and $\vartheta$ the mixing angle in vacuum. For our estimates, we use the simple values $\delta m^2=(50~{\rm meV})^2$ and $\cos2\vartheta=1$. We employ natural units with $\hbar=c=1$.

The DLN is dominated by $\nu_e$, whereas the growth rate is proportional to the small fraction of flipped neutrinos carrying opposite lepton number. This result, derived in Ref.~\cite{Fiorillo:2025zio} and reviewed in the SM~\cite{supplementalmaterial}, reveals that flavomons moving in a direction $\cos\theta$ grow approximately 
with the rate
\begin{equation}\label{eq:growth_rate_narrow}
    \kern-0.3em\gamma(\cos\theta)=\mathrm{max}_{|\bK|}\gamma_\bK\sim \frac{1}{\langle \omega_E^{-1}\rangle(\cos\theta)} \frac{dn_{\rm flip}/d\cos\theta}{ dn_{\rm DLN}/d\cos\theta}.
\end{equation}
Here $n_{\rm flip}$ is the number density of flipped neutrinos, those shaded in red in Fig.~\ref{fig:crossing}, whereas the average is taken over the DLN distribution in the form
\begin{equation}
    \langle \omega_E^{-1}\rangle(\cos\theta)=\frac{\int_{-\infty}^{+\infty} dE \frac{dn_{\rm DLN}}{dEd\cos\theta}\,\omega_E^{-1}}{\frac{dn_{\rm DLN}}{d\cos\theta}}.
\end{equation}

When the number of flipped neutrinos along a given direction reaches the total DLN along that direction, this estimate breaks down, transitioning into a \textit{broad slow instability}~\cite{Fiorillo:2024pns,Fiorillo:2025zio}. The flavomon energy $\Omega_\bK$ has a width $\gamma_\bK\gg\omega_E$ and
\begin{equation}\label{eq:growth_rate_broad}
    \gamma(\cos\theta)=\mathrm{max}_{|\bK|}\gamma_\bK\sim \langle \omega_E\rangle(\cos\theta)
\end{equation}
becomes the maximum growth rate in that direction---the neutrino beams can be approximated as monochromatic.

For very small DLN, the growth rate is enhanced by the small denominator $dn_{\rm DLN}/d\cos\theta$ appearing in the average $\langle \omega_E\rangle (\cos\theta)$. The instability then enters a \textit{non-resonant} regime, where it loses its directional character, i.e., neutrinos with any direction
emit flavomons.
This happens when $\gamma$ becomes comparable with the refractive energy scale $\sqrt{2}\GF n_{\rm DLN}$. (We previously identified this scale as $\mu \epsilon$, where $\mu=\sqrt{2}\GF n_\nu$ and $\epsilon=n_{\rm DLN}/n_\nu$.) This regime, in which the growth rate scales as $\gamma\sim \sqrt{\omega_E \mu}$, never appears in SNe
as it requires a very small DLN.

Our overall estimate of the slow-mode growth rates thus relies on Eqs.~\eqref{eq:growth_rate_narrow} and~\eqref{eq:growth_rate_broad}. Crucially, we associate a growth rate to each neutrino direction $\cos\theta$, which is physically interpreted as the growth rate of flavor waves moving in that particular direction.

{\bf\textit{Spectral crossings in SNe.}}---To assess the impact of spectral crossings, we consider the fraction of flipped neutrinos, which actively emit flavomons. Angular crossings rely on a delicate balance of flavor-dependent angle distributions and depend sensitively on details and dimensionality of the simulations. In contrast, the dominance of $\overline{\nu}_e$ over $\overline{\nu}_\mu$ along certain directions is generic~\cite{Janka:1995cu, Raffelt:2001kv} due to the smaller, hotter decoupling region of $\overline{\nu}_\mu$. Additional effects, such as energy degradation via nucleon scattering and positive temperature gradients in accreting models, reinforce this feature. Hence, the details of the SN model do not significantly affect our results.

We use a numerical model with available angle and energy distributions~\cite{Serpico:2011ir}, namely a spherically symmetric SN simulation of the Garching group, computed with {\sc Prometheus-Vertex}. This code includes energy-dependent three-species neutrino transport with relativistic corrections and state-of-the-art microphysics. The progenitor is a $13.8\,M_\odot$ model of Refs.~\cite{Woosley:2002zz, Woosley:1995ip}, the EoS the one of Lattimer \& Swesty with incompressibility of 180~MeV \cite{Lattimer:1991NuPhA.535..331L}, though results with 220~MeV are nearly identical during the first $100$~ms post-bounce. The model does not explode, which is irrelevant at a few tens of ms. Details of the model are available upon request at the Garching Core-Collapse Supernova Archive~\cite{JankaWeb}.

Spherical symmetry is also not a limiting assumption, as 3D effects are small when slow instabilities develop at some 10\,ms after bounce. Their properties are governed by generic outflow features---notably the growing dominance of $\overline{\nu}_e$ over $\overline{\nu}_\mu$ caused by the shock-heated PNS mantle rapidly growing by high post-bounce mass accretion rates---justifying the use of a single representative model.

\begin{figure*}
\includegraphics[width=\textwidth]{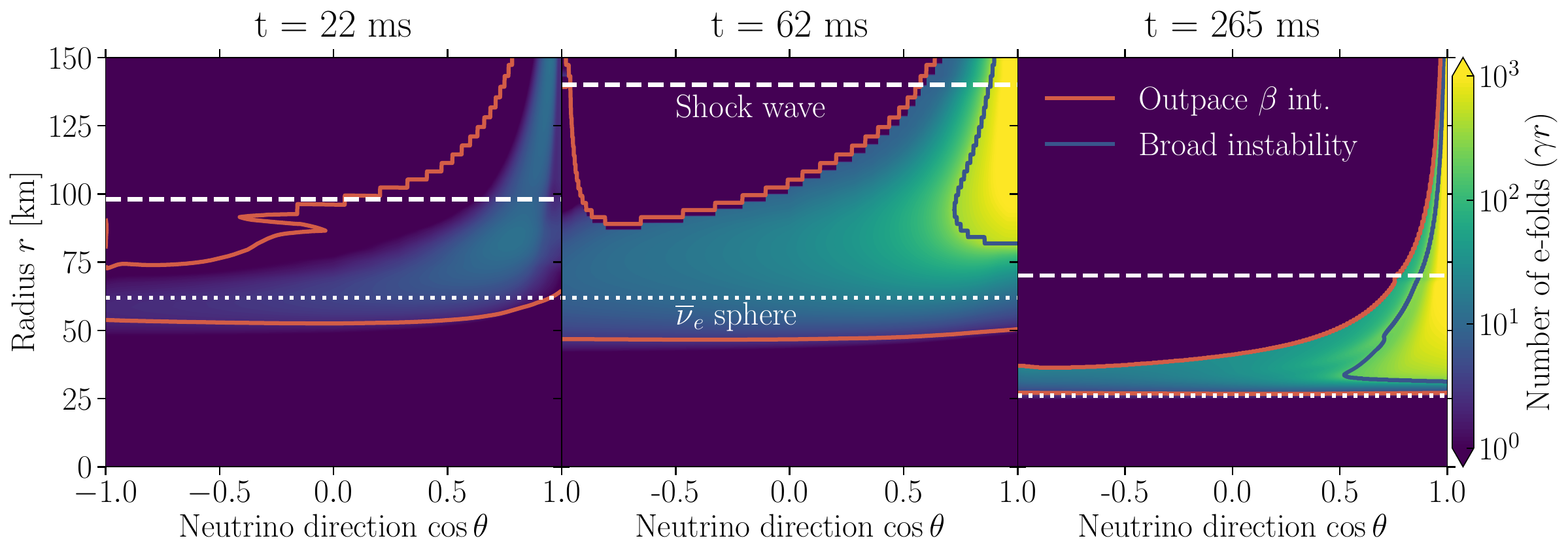}
\caption{Number of e-folds induced by the growth of slow instabilities as a function of neutrino location and direction. The panel layout and 
the positions of neutrinospheres and shock waves are the same as in Fig.~\ref{fig:crossing}.}\label{fig:growth}
\end{figure*}

We first test whether a spectral crossing exists at all, both of the $\overline{\nu}$ and $\nu$ type. For each time, radius, and neutrino direction, we determine the total number of flipped $\overline{\nu}$ (number of $\overline{\nu}_e$ in excess of $\overline{\nu}_x$ in every energy bin) and $\nu$ (same for $\nu_\mu$ in excess of $\nu_e$). The concrete definition, e.g., for $\overline\nu$, is the ratio between the red-shaded area for negative energies in Fig.~\ref{fig:crossing}, and the energy-integrated DLN.
In Fig.~\ref{fig:flipped}, we highlight with a red (blue) contour the regions containing a spectral crossing. 

Crossings of $\overline{\nu}$ appear immediately and over extended regions. They arise first because, while the $\overline{\nu}_e$ and $\overline{\nu}_\mu$ fluxes are comparable, the $\overline{\nu}_\mu$ are more forward directed, causing a $\overline{\nu}_e$ excess in non-forward directions already within the $\overline{\nu}_e$ sphere. With time, the overall $\overline{\nu}_e$ flux dominates over $\overline{\nu}_\mu$ and the crossing extends even to forward directions. Meanwhile, a $\nu$ crossing, corresponding to a high-energy excess of $\nu_\mu$ over $\nu_e$, becomes progressively more significant in the forward direction. However, it is less prominent and occurs after $\overline{\nu}$ crossings, which thus could affect the appearance of $\nu$ crossings.

Since $\overline{\nu}$ crossings dominate, we show the fraction of flipped $\overline{\nu}_e$ in Fig.~\ref{fig:flipped} by color coding, providing a direct handle on the importance of CFCs triggered by spectral crossings; this is defined as the integral of the red-shaded area for negative energies in Fig.~\ref{fig:crossing}, divided by the energy-integrated number of $\overline{\nu}_e$. Already at early times, up to 40\% of anti-neutrinos along non-forward directions are flipped and will emit flavomons, and this fraction strongly increases with time, even reaching up to 100\%. These spectral crossings develop well below the shock wave, suggesting a potential impact on the explosion.

{\bf\textit{Dynamical impact.}}---Spectral crossings alone are not enough to assure a dynamical impact. We need to demonstrate that the resulting instabilities are rapid enough to (i)~outpace the rate of beta interactions, which compete with the decay $\overline{\nu}_e\to \overline{\nu}_\mu+\psi$ in pinning neutrinos to their flavor states, and (ii)~grow substantially to become nonlinear over the dynamical timescale $t_{\rm dyn}\sim r$ for neutrino propagation through the SN. This estimate for $t_{\rm dyn}$ is only approximate because flavomons propagate through a radial profile with a space-changing growth rate. For an order-of-magnitude estimate, we may still assume that there is a single relevant length scale, namely the radius itself. A full study of the propagation through a stratified environment is beyond our scope; more comments are provided in the SM~\cite{supplementalmaterial}.

We estimate the growth rate from the smallest among Eqs.~\eqref{eq:growth_rate_narrow} and~\eqref{eq:growth_rate_broad}, determining which class is relevant.
To assess the dynamical impact, 
we use the number of e-folds $\gamma r$ developed over the dynamical timescale. When it strongly exceeds unity, the instability develops fully. To estimate the role of collisions, we compare the growth rate with the energy-averaged $\nu_e n\to e^- p$ rate, taken from from Eq.~(11) of Ref.~\cite{Janka:2000bt}.

Figure~\ref{fig:growth} shows the $\gamma r$ (e-folds) distribution. Initially non-forward directions dominate the instability, as the forward-peaked $\overline{\nu}_x$ suppress spectral crossings. Later, $\overline{\nu}_e$ dominate across the entire angular range. As time progresses, the instability strengthens with the increasing population of flipped neutrinos, eventually entering the broad regime, within the region marked by the blue line. However,
the growth rate remains below that of fast instabilities and never reaches the non-resonant limit $\gamma \sim \sqrt{\omega_E \mu}$, confirming the findings of Refs.~\cite{Fiorillo:2024pns, Fiorillo:2025ank, Fiorillo:2025zio}. Except for a narrow region below 50~km, the growth rate exceeds that of beta reactions, so collisions do not hinder the instability. 
This illustrates the central point of our work: within tens of ms, slow instabilities grow rapidly enough to alter the flavor distribution.

\textit{\textbf{Discussion.}}---Spectral crossings are generic in SNe. We have shown that $\overline{\nu}$ crossings, caused by $\overline{\nu}_e$ dominating over $\overline{\nu}_\mu$, induce instabilities rapid enough to alter flavor evolution within tens of ms after bounce. The affected region starts below the $\overline{\nu}_e$ sphere and reaches beyond the shock wave. The impact on shock revival is difficult to assess without understanding the nonlinear feedback on the medium. At early times, the instability is driven by flipped $\overline{\nu}_e$, so CFCs will attempt to remove them via $\overline{\nu}_e\to \overline{\nu}_\mu+\psi$, 
suggesting negative feedback on the explosion by reducing the rate of energy deposition. Later effects such as $\nu$ crossings, which would tend in the opposite direction by converting high-energy $\nu_\mu\to \nu_e+\psi$, cannot be captured consistently by post processing. Most importantly, at the early times of interest here, no angular crossing develops~\cite{Nagakura:2021hyb}, so there are no fast instabilities.

While we use a specific numerical SN model for convenience, our argument for the early emergence of slow instabilities is 
generic. $\overline{\nu}$ crossings always lead to instability~\cite{Fiorillo:2025kko}, and they only require a directional dominance of $\overline{\nu}_e$ over $\overline{\nu}_\mu$. In the core, such crossing is precluded by chemical equilibrium, with $\overline{\nu}_e$ suppressed by a negative chemical potential, but outside of the core $\overline{\nu}$ crossings are induced by the luminosity of accretion-powered $\overline{\nu}_e$ overtaking $\overline{\nu}_\mu$ (see, e.g., Fig.~2 of Ref.~\cite{Fiorillo:2023frv}).

These slow instabilities are quite insidious: their wavelengths are even millimeters at these early times, they oscillate over picoseconds, yet grow only over tens of microseconds~\cite{Fiorillo:2024pns, Fiorillo:2025ank, Fiorillo:2025zio}. This is quick enough to alter flavor evolution in SNe, yet slow compared with their rapid oscillations as to obstruct numerical analysis. Conversely, they grow so slowly that their initial amplitude is another concern, a question deferred to the SM~\cite{supplementalmaterial}. Even worse, their initial growth timescale is comparable with the dynamical timescale of escaping neutrinos, so their evolution is nonlocal. Flavomons themselves have a preferred direction along the streaming velocity, initially slightly nonforward, and later entirely in the forward direction. A possible way forward is to treat flavomons as independent degrees of freedom, with their own streaming velocity, affected by the inhomogeneity of the medium~\cite{Fiorillo:2024qbl,Fiorillo:2025ank,Fiorillo:2025npi}.

It is well known, of course, that neutrino-mass-driven instabilities may appear without angular crossings. Ref.~\cite{DedinNeto:2023ykt} noted that close to an angular crossing, slow instabilities might appear---their reference cases show broad or non-resonant ones---and Ref.~\cite{Shalgar:2024gjt} has identified such conditions at about 50~ms. Our configurations, however, are far from an angular crossing and mainly produce narrow slow instabilities. With mm to cm scales, one can find them only with our analytic tools. Beyond identifying these early slow instabilities, we highlight their implications: the induced flavor conversions jeopardize the later appearance of angular crossings, or at least any robust conclusions about their existence.

In the innermost region (below 50\,km), spectral crossings occur where beta reactions are fast. Collisions may here suppress the instability or trigger a collisional instability \cite{Johns:2021qby} (CFI). However, this requires not only a spectral crossing, but also a sufficiently large $\nu_e$–$\overline{\nu}_e$ asymmetry. An approximate condition is $n_{\nu_e}/\Gamma_{\nu_e} \lesssim n_{\overline{\nu}_e}/\Gamma_{\overline{\nu}_e}$, where $\Gamma_\nu$ the beta reaction rate \cite{Fiorillo:2025zio}. Since $\Gamma_{\nu_e}/\Gamma_{\overline{\nu}_e} \sim n_n/n_p$, this implies $n_{\overline{\nu}_e} \gtrsim n_{\nu_e} n_p / n_n$, with $n_n$ and $n_p$ the neutron and proton densities and Pauli blocking was neglected. This condition is only marginally satisfied in our simulation, mostly in regions where the slow instability growth rate is much larger than the beta rate. 

Hence, since CFI have a threshold, unlike slow instabilities, they are generically less important. This conclusion is strongly vindicated by a recent study~\cite{Wang:2025vbx}, showing that the typical growth rate of CFIs is $\gamma\sim 100\,\mathrm{s}^{-1}$, generally smaller than the ones of slow instabilities in our work, which start at $\gamma\sim 10^{-1}\,\mathrm{km}^{-1}\sim 10^4\,\mathrm{s}^{-1}$ at 20~ms, but rapidly grow larger by several orders of magnitude as the crossing gets deeper. At a few 100~ms, CFIs appear in the PNS convective region with larger growth rates due to the small $\nu_e$–$\overline{\nu}_e$ asymmetry, which however is long after the first appearance of slow instabilities.


We finally stress again that searching for instabilities by post processing numerical simulations is a purely diagnostic tool. At best, the only realistic instabilities are the first ones to appear. Figure~\ref{fig:growth} illustrates the progressive breakdown of post-processing. At early times, instabilities are found in a narrow region with small growth rate, whereas later, the number of e-folds becomes progressively larger, signaling a complete loss of realism. This picture agrees with the message highlighted in Refs.~\cite{Fiorillo:2024bzm, Fiorillo:2024uki, Fiorillo:2024pns, Fiorillo:2025ank, Fiorillo:2025npi, Fiorillo:2025zio} that weak instabilities are the only ones that can be consistently inferred from post-processing. Already at about 20~ms, CFCs will prevent the emergence of strong instabilities. Moreover, since these slow instabilities appear already at tens of ms, long before angular crossings~\cite{Nagakura:2021hyb}, the impact of the latter is directly questioned. 

A similar reexamination of what are the first and therefore most important instabilities is also needed for neutron-star mergers, where post-processing reveals ubiquitous fast instabilities (e.g. Refs.~\cite{Wu:2017qpc,Wu:2017drk,Li:2021vqj,Nagakura:2025hss}). We plan to address this question in future.

{\bf\textit{Acknowledgments.}}---We acknowledge Sajad Abbar for sharing important insights, Manuel Goimil-Garc{\'\i}a, Luke Johns, and Hiroki Nagakura for comments on the manuscript, and Lorenz H\"udepohl for preparing the model data including neutrino intensities from his PhD research for the Garching Core-Collapse Supernova Data Archive. DFGF is supported by the Ale\-xander von Humboldt Foundation (Germany), whereas in Munich, we acknowledge partial support by the German Research Foundation (DFG) through the Collaborative Research Centre ``Neutrinos and Dark Matter in Astro- and Particle Physics (NDM),'' Grant SFB--1258--283604770, and under Germany’s Excellence Strategy through the Cluster of Excellence ORIGINS EXC--2094--390783311.

\bibliographystyle{bibi}
\bibliography{References}

\onecolumngrid

\onecolumngrid
\appendix

\setcounter{equation}{0}
\setcounter{figure}{0}
\setcounter{table}{0}
\setcounter{page}{1}
\makeatletter
\renewcommand{\theequation}{S\arabic{equation}}
\renewcommand{\thefigure}{S\arabic{figure}}
\renewcommand{\thepage}{S\arabic{page}}

\begin{center}
\textbf{\large Supplemental Material for the Letter\\[0.5ex]
\mbox{Neutrino-Mass-Driven Instabilities as the Earliest Flavor Conversion in Supernovae}}
\end{center}
\bigskip

In this Supplemental Material, we first review in Sect.~A the theory of flavor kinetic instabilities that was recently developed in a long series of papers, meant both as a guide to the results and to the pertinent literature. Section~B reports properties of the slow instabilities in our SN model for more time snapshots than shown in the main text.

\bigskip

\twocolumngrid

\section{A.~Flavor instabilities in the neutrino plasma}

In this section, we review the theory of flavor kinetic instabilities that was recently developed from first principles in analogy to the theory of plasma kinetic instabilities \cite{Fiorillo:2023mze,Fiorillo:2023hlk, Fiorillo:2024fnl, Fiorillo:2024qbl, Fiorillo:2024bzm, Fiorillo:2024uki, Fiorillo:2024pns, Fiorillo:2024dik, Fiorillo:2025ank, Fiorillo:2025npi, Fiorillo:2025zio}. This approach is centered on the idea of flavor waves representing an independent degree of freedom. We provide a compact summary of the results, as well as guiding the reader through what is found in which of these different papers.

A neutrino plasma can sustain collective flavor oscillations (flavor waves) that are analogous to density waves in ordinary plasmas. These are coherent modulations in flavor content with fixed wavelength and frequency. The quanta of such waves, dubbed flavomons~\cite{Fiorillo:2025npi}, are conceptually similar to plasmons or phonons. 

In a two-flavor ($\nu_e,\nu_\mu$) system, a flavomon $\psi$ carries one unit of $\mu$-lepton number and minus one of $e$-lepton number, corresponding to a coherent $\nu_e \leftrightarrow \nu_\mu$ oscillation. One may view $\psi$ as a particle--hole pair, with $\nu_\mu$ as the particle and $\nu_e$ as the hole. For three flavors, there are three species of flavomons, corresponding to the off-diagonal SU(3) generators $\overline{e}\mu$, $\overline{e}\tau$, and $\overline{\mu}\tau$, and complex conjugates~\cite{Fiorillo:2025npi}. The on-diagonal SU(3) generators correspond to anticorrelated density fluctuations of the different flavors, without off-diagonal component, and were thus dubbed neutrino--plasmons~\cite{Fiorillo:2025npi}. They are essential in the nonlinear interactions of flavomons, but for weak instabilities they are probably unimportant.

Neutrinos can spontaneously emit flavomons, e.g., via $\nu_\mu \to \nu_e + \psi$, thereby converting flavor and transferring lepton number. We here assume an excess of electron flavor, as typical in SNe. If there is a local $\nu_\mu$ or $\overline{\nu}_e$ excess in a given energy or angular range, these states can shed their ``wrong'' lepton number through decays such as $\nu_\mu \to \nu_e + \psi$ or $\overline{\nu}_e \to \overline{\nu}_\mu + \psi$. The kinematics of these processes depend on the relevant energy scales.

The overall weak-potential scale for neutrino-neutrino refraction is $\mu = \sqrt{2} \GF n_\nu$, set by the neutrino density $n_\nu$. However, since the flavomon represents a flavor oscillation, its typical energy is set by the difference ($e-\mu$) lepton number (DLN) introduced in the main text. The DLN ratio in comparison with the total number of neutrinos is measured by the asymmetry
\begin{equation}
\epsilon = \frac{n_{\rm DLN}}{n_\nu}=\frac{n_{\nu_e}-n_{\nu_\mu}-n_{\overline{\nu}_e}+n_{\overline{\nu}_\mu}}{n_\nu},
\end{equation}
which characterizes the global flavor imbalance. In terms of this parameter, flavomons carry energy and momentum of order
\begin{equation}
\Omega,\bK \sim \mu \epsilon.
\end{equation}
Flavomon emission changes the energy of the parent neutrino by an amount $\sim \mu\epsilon$, an effect first identified in Ref.~\cite{Fiorillo:2024fnl} by treating flavomons as classical waves, and later interpreted as decay kinematics in Ref.~\cite{Fiorillo:2025npi}. In SN environments, $\mu\epsilon \ll E$ always holds, i.e., typical neutrino energies far exceed  their refractive interaction energy.

While $\epsilon$ describes the global asymmetry, certain instabilities (notably slow ones) are governed by the local, directional asymmetry. If $\bn$ denotes a propagation direction, the directional asymmetry $\epsilon(\bn)$ is defined analogously, but using angle-differential distributions in place of angle-integrated ones.

The other relevant scale is the vacuum energy splitting between $\nu_e$ and $\nu_\mu$ caused by neutrino masses. While in vacuum, the key effect of masses is flavor mixing, in SNe the large matter potential suppresses oscillations, making the energy splitting (not flavor mixing) the dominant effect of the neutrino mass matrix. This splitting is given by $\omega_E = \delta m^2 \cos 2\vartheta / 2E$, where $\delta m^2$ is the squared-mass difference and $\vartheta$ the mixing angle. (We use the symbol $\theta$ for the direction of motion.) Notice that $\cos2\vartheta$, not $\sin2\vartheta$, governs instabilities, since it is the energy shift (not the off-diagonal mixing) that can trigger them. A detailed discussion of this point, including the role of matter, is provided in Ref.~\cite{Fiorillo:2024pns}. 

In SN environments of interest, the relevant scales are usually ordered according to the hierarchy
\begin{equation}
    \omega_E \ll \mu \epsilon \ll E,
\end{equation}
reflecting the fact that neutrinos are ultra-relativistic particles, and that flavomon energies are much smaller than neutrino energies, yet typically larger than the vacuum energy splitting.

The basic emission process $\nu_\mu \to \nu_e + \psi$ must conserve energy. The initial $\nu_\mu$ energy is
\begin{equation}
    E_{\nu_\mu}=E(\bp)+\frac{\omega_E-\Delta}{2},
\end{equation}
where $E(\bp)=|\bp|$ is the kinetic energy of a massless neutrino. $\Delta$ is the weak interaction potential produced by neutrinos, which we write as $\Delta=\mu(D_0-\bD_1\cdot \bv)$, where $D_0=\epsilon$ is the energy- and angle-integrated DLN, $\bD_1$ the corresponding flux, and $\bv=\partial E/\partial \bp$ the neutrino velocity. The final $\nu_e$, after emitting a flavomon with momentum $\bK$, has $E_{\nu_e} = E(\bp - \bK) - (\omega_E-\Delta)/2 \simeq E(\bp) - \bv \cdot \bK - (\omega_E-\Delta)/2$. Therefore, introducing the shifted frequency $\omega=\Omega+\mu D_0$ and shifted wavevector $\bk=\bK+\mu \bD_1$, energy conservation yields
\begin{equation}\label{eq:energy_conservation}
\omega(\bk) - \bv \cdot \bk - \omega_E = 0,
\end{equation}
where for antineutrinos, $\omega_E \to -\omega_E$. Condition~\eqref{eq:energy_conservation} can be viewed either as a resonance for Cherenkov-like emission of flavor waves~\cite{Fiorillo:2024bzm, Fiorillo:2024uki, Fiorillo:2024pns}, or as the kinematic condition for decay~\cite{Fiorillo:2025npi}. When the flavomon is unstable, $\omega$ acquires an imaginary part $\gamma$, allowing a small violation of Eq.~\eqref{eq:energy_conservation} by terms of order $\gamma$.

If there are no DLN crossings, meaning that for all energies and directions the same flavor $e$ or $\mu$ dominates, lepton number conservation prevents an instability~\cite{Fiorillo:2024bzm}. For fast instabilities, this connection was first established in Ref.~\cite{Johns:2024bob} and was also formally proven using the algebraic properties of the flavor matrix~\cite{Morinaga:2021vmc, Dasgupta:2021gfs}. 

When a crossing first appears in the SN evolution, it must initially emerge in the form of only few neutrinos in an energy and angular range with opposite lepton number to the dominant one. These are the \textit{flipped neutrinos} first introduced in Ref.~\cite{Fiorillo:2024bzm}. They spawn the instability by decaying into flavomons. The concept of flipped neutrinos is rigorous only for weak instabilities, when the crossing has just appeared. As stressed in Refs.~\cite{Fiorillo:2024bzm, Fiorillo:2024uki}, this is the \textit{only} realistic case because, as soon as a crossing emerges, flavor conversions will begin and alter the subsequent evolution. This likely holds even as the instability develops nonlinearly, since it will tend to remove the cause that led to its appearance. The evolution would then proceed at the edge of instability, as shown in Ref.~\cite{Fiorillo:2024qbl} for fast instabilities.

With this understanding, we can next categorize the different regimes of instability. Our classification is based on comparing the typical growth rate $\gamma$ of flavomon emission with the energy spread among neutrinos of different energies and directions. The main regimes are:
\begin{itemize}
\item \textbf{Fast (angular) instability:} An angular crossing, i.e., a sign change of the energy-integrated DLN across directions, triggers a fast instability. A typical growth rate is
\begin{equation}
    \gamma \sim \mu \left( \frac{n_{\rm flip}}{n_\nu} \right),
\end{equation}
proportional to the fraction of flipped neutrinos. A more exact expression in Ref.~\cite{Fiorillo:2024uki} yields excellent agreement with a numerical solution of the dispersion relation. The physically interesting case, when $n_{\rm flip}\ll n_\nu \epsilon$ and the instability is weak, has $\gamma \ll \omega$, so flavomons are emitted nearly on-shell. Emission is then primarily collinear with the flipped neutrinos, whereas along the direction of non-flipped neutrinos, $\nu_e+\psi\to \nu_\mu$ always dominates over $\overline{\nu}_e\to \overline{\nu}_\mu+\psi$ and $\nu_\mu\to \nu_e+\psi$. The instability is therefore convective~\cite{Fiorillo:2024dik, Fiorillo:2025ank}, implying that flavomons stream away from their source region along the direction of the angular crossing.

\item \textbf{Slow (mass driven) instability:} If there is no angular crossing, an instability can still be driven by a spectral crossing, either in the angle-integrated distribution, or along specific directions. The subsequent evolution depends on how large is $\gamma$ compared with $\omega_E$ and with the velocity-dependent part of the change in the neutrino energy $\bk\cdot \bv\sim \mu \epsilon$. We identify three regimes, in order from the weakest to the strongest instability:
\begin{itemize}
    \item \textbf{Narrow instability:} For shallow spectral crossings along a direction $\bn$, the instability is weak and has a small growth rate. Only flipped neutrinos along $\bn$ can emit flavomons, and only due to their $\omega_E$. If this energy splitting would vanish, the amplitudes from different energies would cancel, preventing the instability. An analytical approximation for $\gamma$ \cite{Fiorillo:2025zio} perfectly matches the numerical solution in this regime. The maximum growth rate for flavomons along $\bn$ is approximately
\begin{equation}
\gamma \sim \frac{\pi n_{\rm flip}(\bn)}{\int_{-\infty}^{+\infty} dE\, \omega_E^{-1} dn_{\rm DLN}(\bn)/dE},
\end{equation}
where for compactness we denote by $n_\nu(\bn)$ the differential neutrino distribution per unit solid angle. This expression is more easily interpreted after introducing the average over the DLN energy distribution, as done in the main text. Neglecting the factor $\pi$ in the spirit of an order-of-magnitude estimate, we recover the expression used in the main text
\begin{equation}
    \gamma\sim \frac{1}{\langle \omega_E^{-1}\rangle(\bn)}\frac{n_{\rm flip}(\bn)}{n_{\rm DLN}(\bn)}
\end{equation}
except for notation. It is intuitive that this expression corresponds to the energy scale of the vacuum frequency $\langle \omega_E^{-1}\rangle(\bn)^{-1}$, suppressed by the fraction of flipped neutrinos, since only they can emit flavomons. Crucially, the instability is \textit{directional}, because the growth rate is too small to allow neutrinos with different directions to contribute to flavomon emission, given the energy cost $\bk\cdot \bv\sim \mu \epsilon$. The main feature of this regime is that $\gamma\ll\omega_E\ll \mu \epsilon$.

\item \textbf{Broad instability:} When $n_{\rm flip}(\bn)$ becomes comparable to $n_{\rm DLN}(\bn)$, the spectral crossing deepens and the growth rate approaches or exceeds the typical $\omega_E$. In this regime, energy conservation [Eq.~\eqref{eq:energy_conservation}] can be violated by $\gamma \gg \omega_E$, allowing neutrinos of any energy to emit flavomons. The growth rate is then set by the energy-integrated DLN, and an approximate explicit expression was derived in Ref.~\cite{Fiorillo:2024pns} for monochromatic and in Ref.~\cite{Fiorillo:2025zio} for continuous energy distributions. In both cases, this approximation is extremely precise compared with a numerical solution. The predicted typical growth rate is
\begin{equation}
    \gamma\sim \langle \omega_E\rangle.
\end{equation}
Notice that, for a gas of $\nu_e$ and $\overline{\nu}_e$, the order of magnitude of this expression is $\gamma\sim \omega_E/\epsilon(\bn)$, i.e., the growth rate is enhanced by a small DLN.
For $\epsilon(\bn)\ll1$, corresponding to a deep crossing, the flavomon energy is distributed over a wide range $\Delta E \sim \gamma \gg \omega_E$. This regime, first identified in Ref.~\cite{Fiorillo:2024pns}, is characterized by $\omega_E\ll \gamma \ll\mu \epsilon$. Flipped neutrinos of all energies can emit flavomons, but only if they are collinear with them, due to the energy cost $\bk\cdot \bv\sim \mu \epsilon\gg \gamma$ for neutrinos with different directions to produce them.

\item \textbf{Non-resonant instability:} When $\epsilon(\bn)$ is so small that $\gamma$ becomes comparable with the flavomon energy $\mu \epsilon$ itself, neutrinos of all directions can contribute to emission, because the flavomon energy is broadened by $\gamma \gtrsim \mu \epsilon$, larger than the energy cost for non-collinear neutrino decay. The instability is now non-resonant, and its typical growth rate is $\gamma\sim \sqrt{\omega_E\mu}$, which can only occur if $\epsilon \ll \sqrt{\omega_E/\mu}$. The slow pendulum instability, identified long ago~\cite{Samuel:1993uw,Hannestad:2006nj}, belongs to this class, although it is a very special case due to its many symmetries~\cite{Fiorillo:2024pns}. This regime is probably not very relevant in practice because the asymmetry between neutrinos and antineutrinos must be extremely small.
\end{itemize}
\end{itemize}

We have always neglected the effect of collisions. The collisional instability \cite{Johns:2021qby} (CFI) is triggered by energy- and flavor-dependent beta-interactions. According to Ref.~\cite{Fiorillo:2025zio}, narrow CFIs do not exist, so they cannot arise when a spectral crossing first appears; they only emerge once the crossing is sufficiently deep. By order of magnitude, the excess of $\overline{\nu}_e$ must be as large as discussed in the main text and then, CFIs can become competitive in regions where their rate exceeds the slow instability rate. However, the results in the main text suggest that slow instabilities appear first and in a very extended region, so they might prevent the CFI development. Generally, collisions occur at rates $\Gamma_E\ll \mu\epsilon$ and can be treated as a separate, slow dynamics, although this has only been explored in simplified homogeneous setups~\cite{Fiorillo:2023ajs}.

Another key issue of the linear flavomon phase is their space-time evolution: where do they go after being born? This question relates to the absolute vs.\ convective nature of the instability~\cite{Capozzi:2017gqd, Yi:2019hrp}. If it is weak, either fast or slow, flavomons emerge collinearly with flipped neutrinos~\cite{Fiorillo:2024bzm, Fiorillo:2024uki, Fiorillo:2024pns}, propagating away from their source and producing convective growth~\cite{Fiorillo:2025ank}. Their group velocity is affected by matter inhomogeneities and satisfies the WKB-like equations:
\begin{equation}
\dot{\br} = \partial_\bK \Omega(\bK, \br)
\quad\mathrm{and}\quad
\dot{\bK} = -\partial_\br \Omega(\bK, \br).
\end{equation}
The description of flavomons as independent degrees of freedom, with their own streaming velocity, was first introduced in Ref.~\cite{Fiorillo:2025npi}. Meanwhile, the concept of geometric-optics description of streaming flavor waves was independently developed \cite{Johns:2025yxa}.  Notice that, since the flavomon frequency $\Omega(\bK, \br)$ is largely dominated by the $\bK$-independent matter refraction term $\lambda=\sqrt{2}\GF n_e$, where $n_e$ is the electron density (e.g.~Ref.~\cite{Fiorillo:2025ank}), a large-scale inhomogeneous matter density causes flavomon drift in momentum space, a topic for future study. 

This drift treatment also shows that the effect of inhomogeneities on the modes will be strong when $|\dot{\bK}|\delta t \sim |\bK|$, where $\delta t\sim \mathrm{Im}(\omega)^{-1}$ is the typical timescale over which modes grow. Since $|\bK|\sim \mu \epsilon$, this implies that inhomogeneities will strongly affect flavomon propagation, and might even hinder instability---by shifting $\bK$ rapidly out of the unstable range---when
\begin{equation}
    |\partial_\br \lambda| \gg \gamma \mu \epsilon.
\end{equation}
This simple estimate is roughly in agreement with the numerical study of a two-beam model in Ref.~\cite{Bhattacharyya:2025gds}, although for the fast instabilities studied there, very large values of $|\partial_\br \lambda|$ are required, likely too large compared with realistic SN environments. For slow instabilities, the effect of inhomogeneities sets in at much smaller gradients, since $\gamma\sim 1$--$10\,\mathrm{km}^{-1}$, so it could affect the slow modes we identify at the smallest radii.

\begin{figure*}
\includegraphics[width=\textwidth]{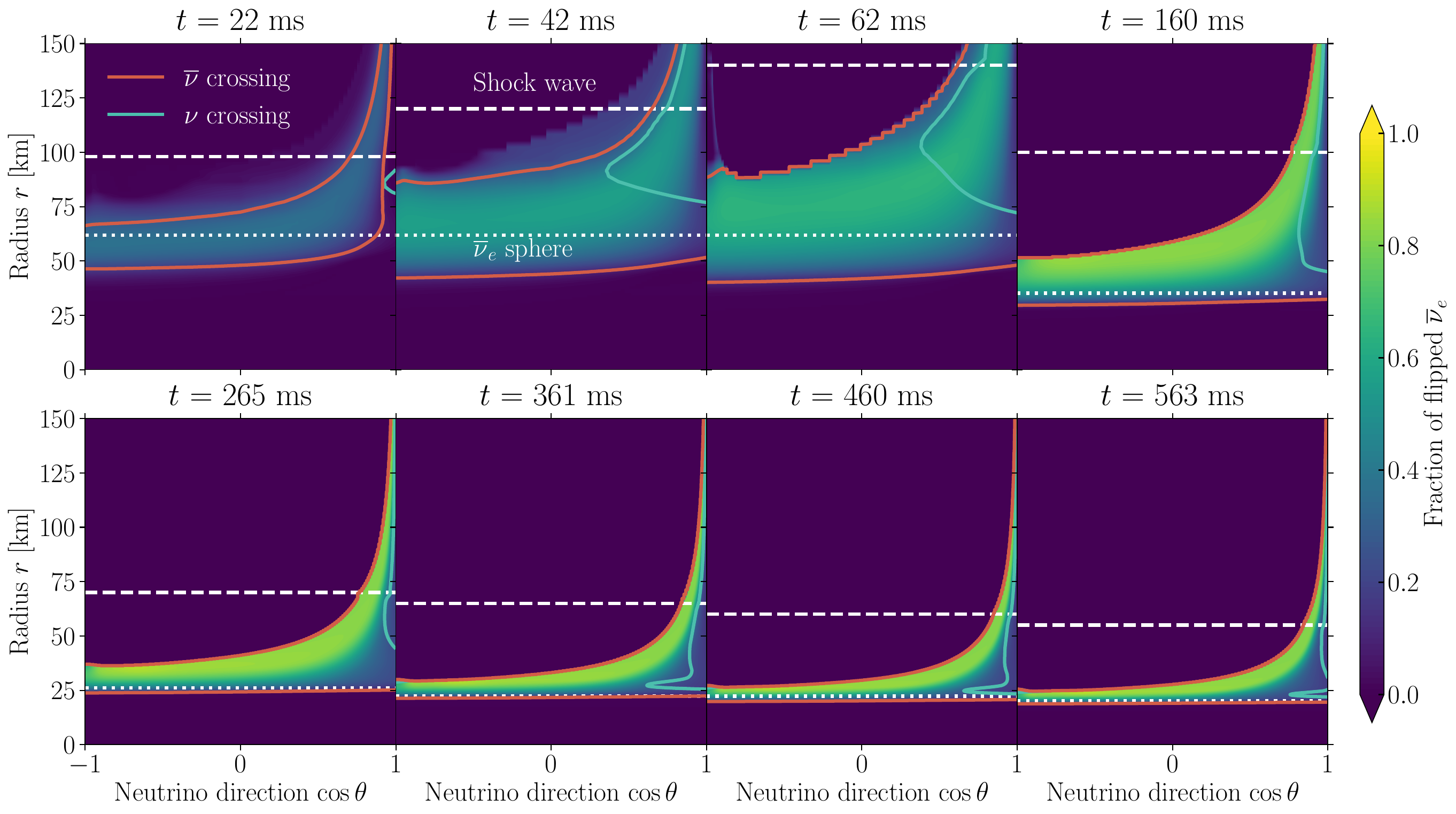}
\caption{Fraction of flipped $\overline{\nu}_e$ as a function of radius and direction. Same as Fig.~\ref{fig:flipped} of the main text, now for all available time snapshots.}\label{fig:flipped_complete}
\vskip12pt
\includegraphics[width=\textwidth]{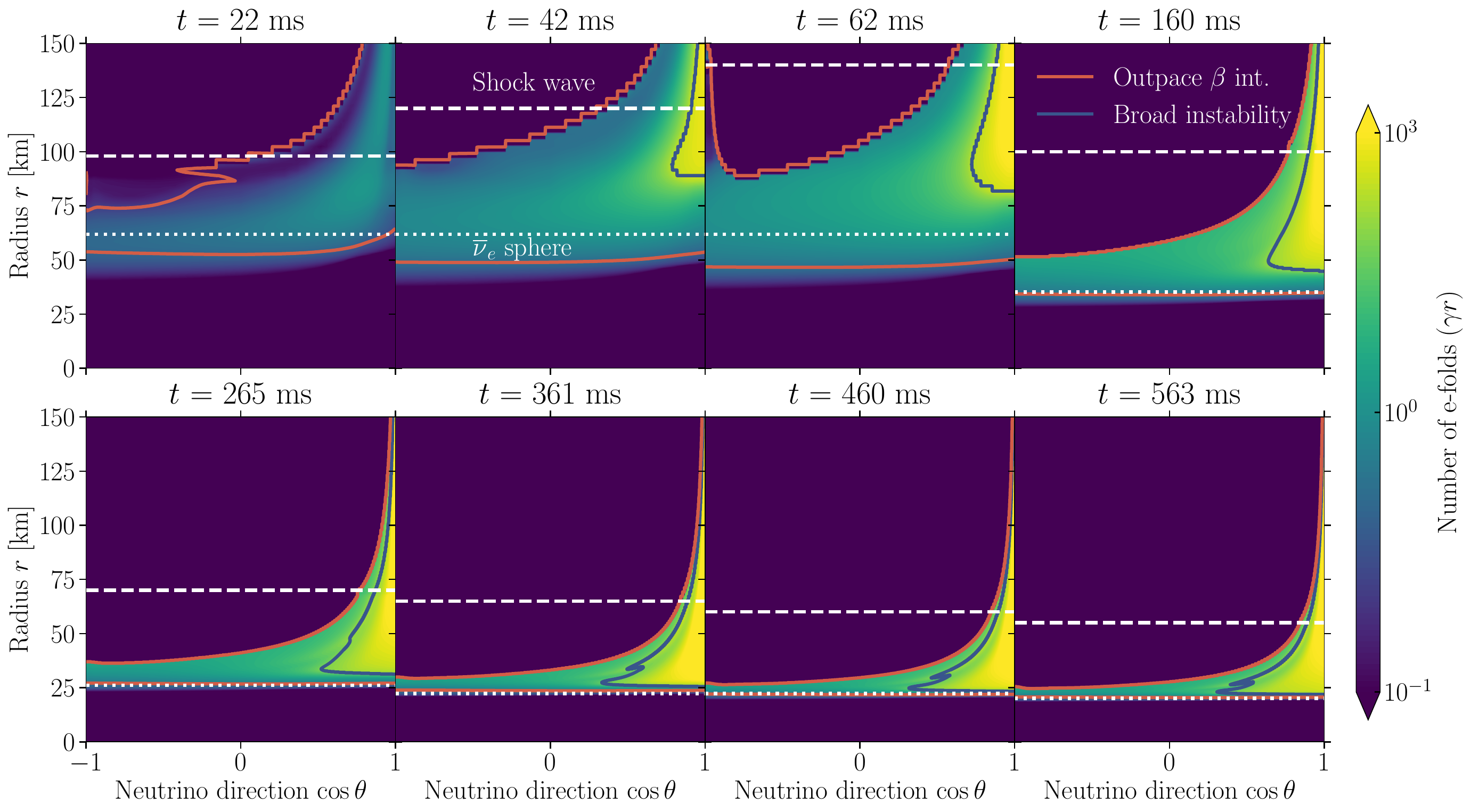}
\caption{Number of e-folds induced by the growth of slow instabilities as a function of neutrino radius and direction. Same as Fig.~\ref{fig:growth} of the main text, now for all available time snapshots.}\label{fig:growth_complete}
\end{figure*}

This concludes our summary of linear theory. The non-linear evolution, when the number of flavomons grows large, is far less understood. In homogeneous systems, the nonlinear regime admits exact solutions, such as the slow and fast pendulum modes~\cite{Samuel:1993uw, Hannestad:2006nj, Padilla-Gay:2021haz, Johns:2019izj}, due to a host of conserved quantities~\cite{Raffelt:2011yb, Pehlivan:2011hp, Fiorillo:2023mze}. These solutions are special cases of flavor solitons~\cite{Fiorillo:2023hlk}, essentially classical solutions of the flavomon field, analogous to standard quantum field theory. However, these solutions rely on symmetries that would quickly break in realistic settings due to instabilities. 

When the system is inhomogeneous, numerical studies (e.g., Refs.~\cite{Bhattacharyya:2020jpj, Nagakura:2022kic, Zaizen:2022cik, Zaizen:2023ihz, Xiong:2023vcm, Cornelius:2023eop, Xiong:2024pue, Abbar:2024ynh, Fiorillo:2024qbl}) have largely focused on local setups dominated by fast instabilities. Yet, since flavomons propagate away from their region of origin, nonlocal effects must arise, casting doubts on the practical relevance of local simulations, as stressed in Ref.~\cite{Fiorillo:2025ank}. 

For slow instabilities, the challenge is even greater: their small growth rates $\gamma \ll \mu\epsilon$ imply that flavomons travel over large distances before becoming nonlinear, making their evolution inherently nonlocal. The only numerical study of inhomogeneous slow instabilities~\cite{Padilla-Gay:2025tko} focused on the nonresonant regime in which $\gamma\gtrsim\mu\epsilon$, so this issue did not show up, but the required asymmetry for this regime appears unrealistically small. Altogether, these considerations point to substantial limitations of a purely numerical approach.

For now, the only predictive theoretical framework for the nonlinear evolution, independent of numerical work, is the quasi-linear theory of neutrino-flavomon interactions \cite{Fiorillo:2024qbl, Fiorillo:2025npi}. It treats flavomons as independent degrees of freedom, neglects their self-interactions, and focuses solely on the neutrino-flavomon coupling.\footnote{Another seemingly different quantitative approach was recently developed, beginning from a 
conceptually different starting point called ``Local-equilibrium theory of neutrino oscillations'' \cite{Johns:2025yxa}, a generalization of the earlier miscidynamics \cite{Johns:2023jjt}. Despite a very different terminology, we recognize a convergence of concepts in that flavor waves are progressively accepted as independent degrees of freedom. While the discussion stays at the formal level, a perturbative expansion is advocated as a way to solve the equations, with the difference that, purely formally, the expansion is around a more general zeroth-order configuration. The only calculational way forward would then be a diagrammatic expansion, leading at linear order to the production of flavor waves, and then, taking into account their backreaction on the original state, to quasi-linear theory. Therefore, we think the broad-brush picture advocated in \cite{Johns:2025yxa} 
also amounts to quasi-linear theory---therein called ``weak inhomogeneity"---once the development is taken beyond the formal level to calculating specific processes. The higher-order terms in the expansion, therein associated with flavor-wave kinetics, correspond to wave-wave coupling, well-known in plasma physics, and introduced as non-linear flavomon vertices in Ref.~\cite{Fiorillo:2025npi}.}
This approach mirrors quasi-linear theory of plasma instabilities \cite{Vedenov1962quasi, drummond1961non, Drummond:1964,schKT}, which remains the primary analytical tool to study their relaxation. In a two-beam case study, it was shown that this theory reproduces local simulation outcomes~\cite{Fiorillo:2024qbl}. Quasi-linear theory leads to the general conclusion that instabilities tend to remove their original cause, leading to an evolution at the edge of instability. The general quasi-linear kinetic equations for neutrino-flavomon interactions were developed in Ref.~\cite{Fiorillo:2025npi}. They bypass the multi-scale nature of flavor instabilities by not resolving the flavomon wavelength, treating its dynamics at linear level. However, a general method for solving these equations remains to be developed.

\section{B.~Complete characterization of the slow instability}

In this section, we report the properties of the slow instabilities, which in the main text were only shown for three representative time snapshots, for all available cases. This will provide a more continuous picture of how the instability evolves from weak to strong, i.e., from a narrow instability with few flipped neutrinos to a broad one with a large flipped population.

\begin{figure*}
\includegraphics[width=0.6\textwidth]{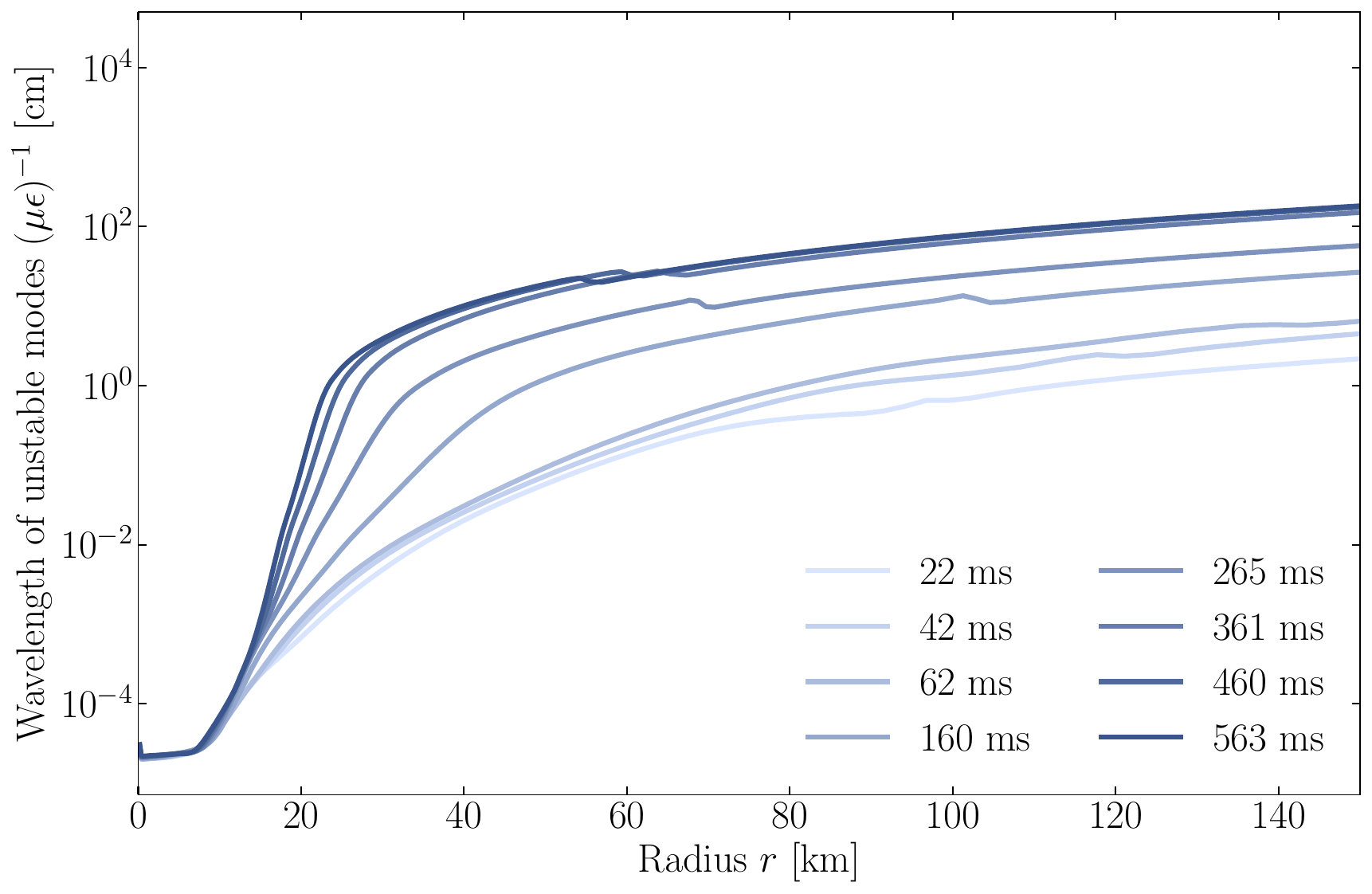}
\caption{Typical wavelength of unstable modes, estimated as $k^{-1}\sim (\mu \epsilon)^{-1}$ \cite{Fiorillo:2024pns, Fiorillo:2025ank, Fiorillo:2025zio}. The progressively darkening color corresponds to increasing time.}\label{fig:wavelength}
\end{figure*}

\begin{figure*}
\vskip12pt
\includegraphics[width=\textwidth]{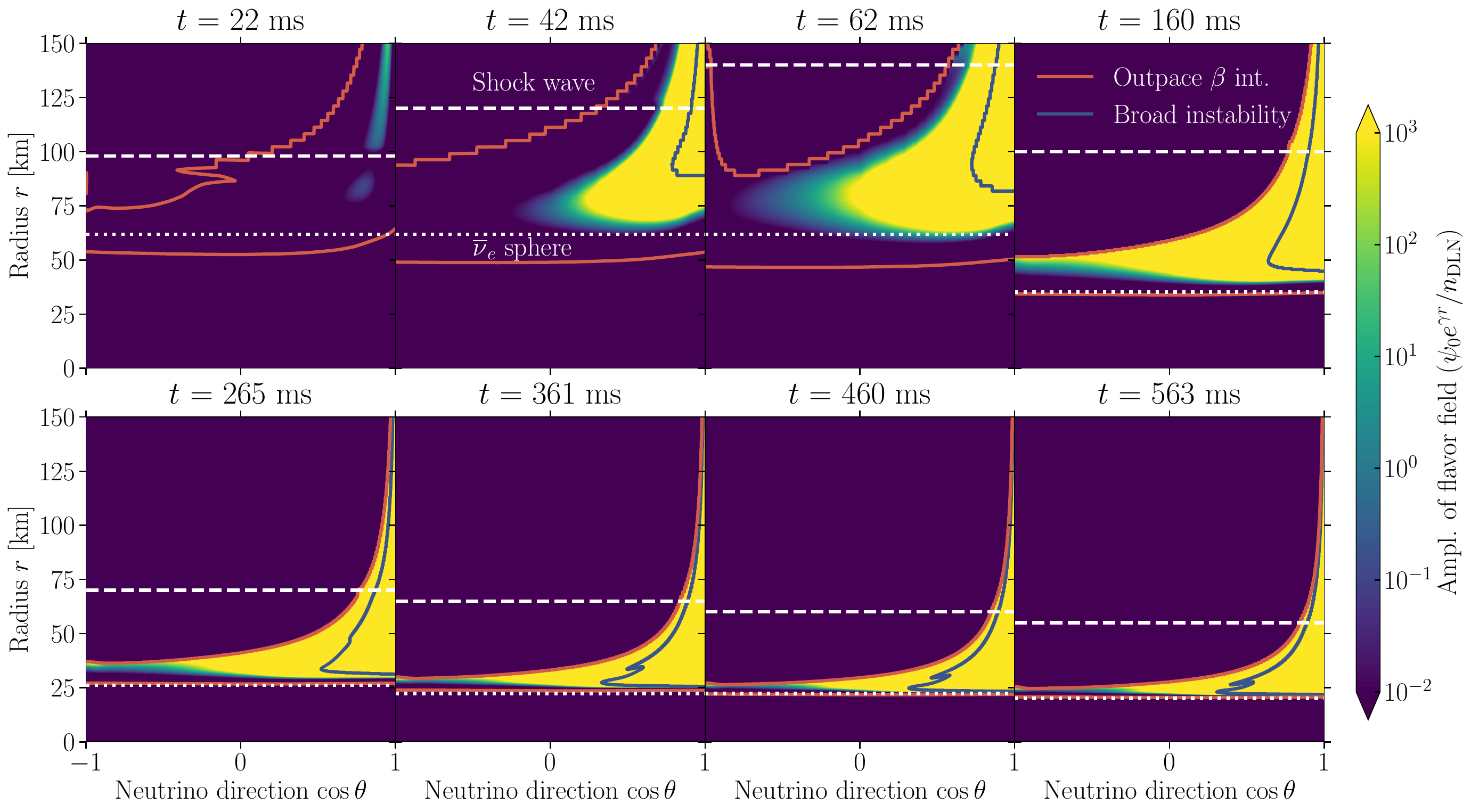}
\caption{Amplitude of the flavor field after the instability has grown through its linear phase, using a natural seeding amplitude as discussed in the text. Figure layout and graphical elements are the same as in Fig.~\ref{fig:growth_complete}. The color legend saturates for an amplitude exceeding $10^3$, but it becomes much larger, showing that even if the seed amplitude $\psi_0$ were overestimated the instability would still saturate very quickly, within $40\,\mathrm{ms}$ at most.}\label{fig:seed}
\end{figure*}

\begin{figure*}
\vskip12pt
\includegraphics[width=\textwidth]{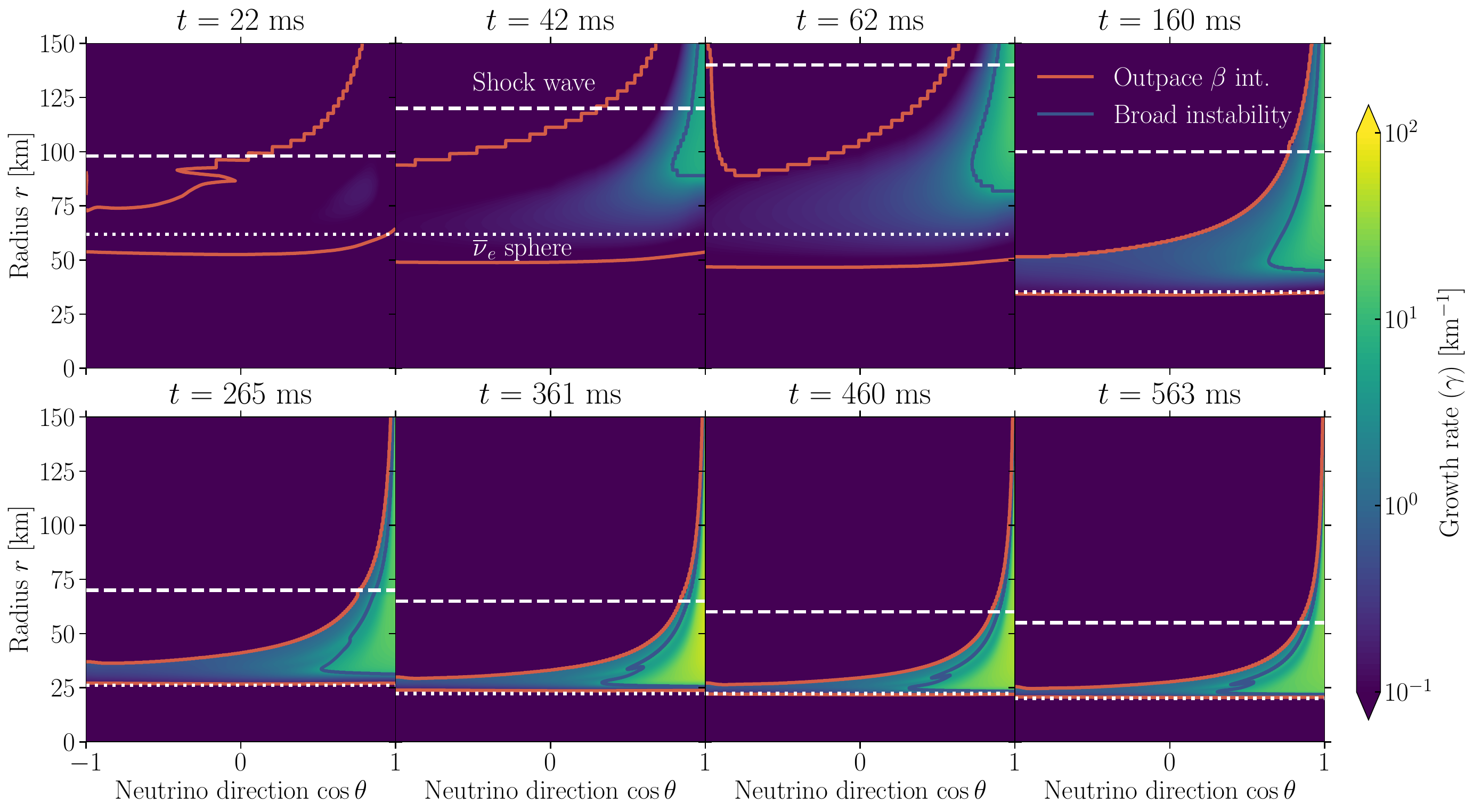}
\caption{Growth rate of slow flavor instability. Figure layout and graphical elements are the same as in Fig.~\ref{fig:growth_complete}.}\label{fig:growth_rate}
\end{figure*}

In analogy to Fig.~\ref{fig:flipped} of the main text, Fig.~\ref{fig:flipped_complete} shows the complete time evolution of the fraction of flipped $\overline{\nu}_e$, together with the regions where a sizable $\overline{\nu}$ and $\nu$ crossing appears. This shows in a more continuous manner the evolution of the instability. Initially, the excess of $\overline{\nu}_e$ is not large, especially in the forward direction,  but it gradually increases and extends to the forward direction. Interestingly, around $60\,\mathrm{ms}$, we also see the formation of a $\overline{\nu}_e$ excess in the backward direction, the so-called halo effect~\cite{Cherry:2012zw, Sarikas:2012vb, Cherry:2013mv, Cirigliano:2018rst, Zaizen:2019ufj}, which however is not strong enough to produce an angular crossing, and therefore does not trigger a fast instability as proposed in Refs.~\cite{Morinaga:2019wsv, Abbar:2021lmm}, yet it is still subject to the slow instability we identify here.

In analogy to Fig.~\ref{fig:growth} of the main text,  Fig.~\ref{fig:growth_complete} shows the number of e-folds for all available snapshots. As in Fig.~\ref{fig:flipped_complete}, it provides a more progressive and continuous picture of the instability evolution. We see that the number of e-folds becomes very large already at about 40~ms, with the development of a $\nu$ crossing close to the shock wave. This picture should really be interpreted as the history of the progressive loss of meaning of post-processing: a number of e-folds much larger than 1 implies full development of the instability and corresponding nonlinear adjustment of the flavor content, rendering the SN model fundamentally inconsistent. 

In addition to the growth rate, the unstable modes are also characterized by their typical wavelengths. As shown in Refs.~\cite{Fiorillo:2024pns, Fiorillo:2025ank, Fiorillo:2025zio}, unstable modes appear only at relatively large wavenumbers, corresponding to a typical wavelength $k^{-1}\sim (\sqrt{2}\GF n_{\rm DLN})^{-1}\sim (\mu \epsilon)^{-1}$, corresponding to the energy- and angle-integrated DLN. It is instructive to see explicitly how large this quantity can be. We show $k^{-1}$ in Fig.~\ref{fig:wavelength} for all available snapshots as a function of radius. The most interesting region is 50--100~km, where the instability develops more strongly at early times as gleaned from Fig.~\ref{fig:growth_complete}. For the earliest snapshots, which are the most relevant ones since they feature the first appearance of the instability, the typical wavelength is between $10^{-2}$ and $1\,\mathrm{cm}$, which is even smaller than expected scales for fast instabilities. Later, with a decrease in the neutrino number density due to the progressively rarefied outflow, the wavelength adjusts to the more typical few-cm scale. Thus, the early appearance of slow instabilities has an additional consequence, namely an even smaller spatial scale of growing modes.

A final point of concern is that, since the growth of the slow modes is not as rapid as the fast ones, the amplitude of the initial flavor field, often called the seed, might be too small to lead to sizable growth. To ease this concern, we now estimate explicitly the amplitude of the exponentially growing flavor field. Initially, neutrinos are in flavor eigenstates, but flavor mixing provides a forcing term. As shown more explicitly in Ref.~\cite{Fiorillo:2024pns}, the resulting amplitude of the flavor field (the off-diagonal component of the neutrino density matrix summed over all momenta) is $\psi\sim \psi_0 e^{\gamma t_{\rm dyn}}$, where $t_{\rm dyn}$ is the dynamical timescale over which the instability evolves. Moreover, the ratio between the frequency of the forcing term and the eigenfrequency of the mode is
\begin{equation}
    \frac{\psi_0}{n_{\rm DLN}}\sim \frac{\delta m^2 \sin 2\vartheta}{2E \Omega_\bK},
\end{equation}
where $\vartheta$ is the mixing angle. For unstable modes, $\Omega_\bK$ is largely dominated by the refractive frequency from electrons, so we can safely approximate $\Omega_\bK\sim \lambda\sim \sqrt{2}\GF n_e$, where $n_e$ is the electron density and $\lambda$ is not to be confused with the wavelength. 

Since we are only interested in the order of magnitude of the resulting amplitude, we can therefore approximate
\begin{equation}\label{eq:initial_seed_amplitude}
    \frac{\psi_0}{n_{\rm DLN}}\sim \frac{\delta m^2}{2E\lambda},
\end{equation}
neglecting the factor $\sin 2\vartheta$, which anyway is not much smaller than 1. Using $t_{\rm dyn}\sim r$, we can thus estimate the amplitude of the flavor field as $\psi/n_{\rm DLN}\sim (\psi_0/n_{\rm DLN}) e^{\gamma r}$ and test whether it becomes comparable to~1. It is clear that, even though $(\psi_0/n_{\rm DLN})$ is generally a very small number, ranging from $10^{-10}$ to $10^{-4}$, when the number of e-folds $\gamma r$ becomes comparable with 20 or larger we will still have a full development of the instability. 

This is indeed what we find. Notice that Eq.~\eqref{eq:initial_seed_amplitude} is somewhat on the optimistic side of the seed amplitude, which is proportional to the amplitude of neutrino density fluctuation at the spatial scales which are unstable. So the seed amplitude can be reduced by the much smaller power of the neutrino density fluctuations at the small length scales of a few centimeters. Nevertheless, since the amplitude grows exponentially with the number of e-folds, this ultimately cannot change our conclusions.

Figure~\ref{fig:seed} shows the resulting amplitude of the flavor field after instability growth. At very early times, around 20~ms, due to the very small amplitude of the mixing term triggering the instability, the flavor field grows sizably only outside of the shock wave. However, already between 20 and 40~ms, the slow instability manages to saturate completely outside of a radius of about 60~km. Therefore, despite the small amplitude of the seeding, suppressed by matter refraction, slow instabilities are still able to affect flavor conversions within a timescale of tens of ms post-bounce.

Finally, for completeness, and to adapt to the more conventional representation, Fig.~\ref{fig:growth_rate} shows the growth rate, rather than the number of e-folds, for all the available time snapshots of the simulation. This quantity has a less direct physical interpretation, as it does not immediately clarify whether the instability is dynamically relevant for the neutrino outflow. Nonetheless, it provides a generic insight into the typical magnitude of the slow instability growth rate, which is directly proportional to the vacuum frequency $\omega_E\sim 0.1-1\,\mathrm{km}^{-1}$, and is enhanced in the broad instability regime by the small DLN in the forward direction (see Eq.~\ref{eq:growth_rate_broad}).

\onecolumngrid


\end{document}